\documentclass[twocolumn,preprintnumbers,superscriptaddress,endnote,nofootinbib,aps,prd]{revtex4-1}

\pdfoutput=1

\usepackage{graphicx}
\usepackage{dcolumn}   % needed for some tables 
\usepackage{amssymb}   % for math
\usepackage{amsmath}
\usepackage[FIGTOPCAP]{subfigure}
\usepackage{float}

\usepackage{colordvi}
\usepackage{epsfig}
\usepackage{latexsym}

\newcommand{ \centeron }[2]{{\setbox0=\hbox{#1}\setbox1=\hbox{#2}\ifdim
                             \wd1>\wd0\kern.5\wd1\kern-.5\wd0\fi \copy0
                             \kern-.5\wd0\kern-.5\wd1\copy1\ifdim\wd0>\wd1
                             \kern.5\wd0\kern-.5\wd1\fi}}
\newcommand{ \ltap }{\>\centeron{\raise.35ex\hbox{$<$}}
                     {\lower.65ex\hbox{$\sim$}}\>}
\newcommand{ \gtap }{\>\centeron{\raise.35ex\hbox{$>$}}
                     {\lower.65ex\hbox{$\sim$}}\>}

\newcommand{ \slashchar }[1]{\setbox0=\hbox{$#1$}   % set a box for #1
   \dimen0=\wd0                                     % and get its size
   \setbox1=\hbox{/} \dimen1=\wd1                   % get size of /
   \ifdim\dimen0>\dimen1                            % #1 is bigger
      \rlap{\hbox to \dimen0{\hfil/\hfil}}          % so center / in box
      #1                                            % and print #1
   \else                                            % / is bigger
      \rlap{\hbox to \dimen1{\hfil$#1$\hfil}}       % so center #1
      /                                             % and print /
   \fi}                                             %

\begin{document}
%\preprint{FERMILAB-PUB-11-xxx-T}

\title{Expected Limits on R-symmetric $\mu \to e $ Processes at Project X}

\author{R. Fok}
\affiliation{Department of Physics and Astronomy, York University, Toronto, ON, Canada, M3J 1P3}

\date{\today}
%%%%%%%%%%%%%%%%%%%%%%%%%%%%%%%%%%%%%%%%%%%%%%%%%%%%%%%%%%%%%%%%%%%%%%%%%%%%

\begin{abstract}
We investigate $\mu \to e$ processes in the Minimal R-symmetric Standrad Model (MRSSM) with the expected limits from Project X. It is found that $\mu \to e$ conversion provides the tightest bound on the $\mu \to e$ mixing parameters at the order of $\lesssim O(10^{-3})$. Whereas $\mu \to eee$ only slightly improves the bound in the region where incoherence among different contributions to $\mu \to e$ is significant. No improvements on the bounds are obtained from $\mu \to e \gamma$.

\end{abstract}

\maketitle

\section{Introduction}
\label{sec:intro}

%\begin{itemize}
%\end{itemize}

Lepton flavor violation (LFV) is predicted to occur at an unobservably 
small rate in the Standard Model (SM).  In low energy supersymmetric
theories, new sources of lepton flavor violation are generic 
in the soft breaking sector.  The experimental non-observation of 
$\mu \rightarrow e$ processes is particularly restrictive, 
given the impressive 
bounds on $\mu \rightarrow e\gamma$ from MEGA \cite{Ahmed:2001eh}
and MEG \cite{Adam:2011ch}; on $\mu \rightarrow e$ conversion
from SINDRUM~II \cite{Bertl:2006up}, and to a lesser extent from 
$\mu \rightarrow 3e$ from SINDRUM \cite{Bellgardt:1987du}.
Further progress is expected from various ongoing experiments as well as planned future experiments such as 
Mu2e \cite{Kutschke:2009zz} and other proposals utilizing
Project X at Fermilab \cite{projectx}.

The slepton mixing parameters are tightly constrained in the MSSM  \cite{Hisano:1995cp,Hisano:1998fj,Masina:2002mv,Paradisi:2005fk,Ciuchini:2007ha} by the above experiments. For instance, the most severe constraint arises from $\mu \to e \gamma$ that involves a left-right slepton mixing. If one parametrizes the mixing as $\delta^l_{XY} =  \delta m^2_{XY} / m^2$, where $\delta m^2_{XY}$ is the $(XY)-$element of the slepton mass squared matrix. Then the non-observation of any $\mu \to e \gamma$ events gives a bound of $\delta^l_{LR} \sim O(10^{-5})$ \cite{Ciuchini:2007ha}.  In models where left-right sfermion mixings are absent, the constraints on sfermion mixing can be significantly relaxed. One can accomplish this by enlarging the $R$-parity in the MSSM to a $U(1)$ continuous symmetry \cite{Fayet:1978qc,Polchinski:1982an,Hall:1990hq}.

$R$-symmetric supersymmetry has inspired recent model building \cite{Davies:2011js, Davies:2011mp,Fox:2002bu,Nelson:2002ca,Chacko:2004mi,Carpenter:2005tz,Antoniadis:2005em,Nomura:2005rj,Antoniadis:2006uj,Kribs:2007ac,Amigo:2008rc,Benakli:2008pg,Blechman:2009if,Carpenter:2010as,Kribs:2010md,Abel:2011dc,Frugiuele:2011mh,Itoyama:2011zi}. The MRSSM features Dirac gauginos and their phenomenology has been extensively studied \cite{Davies:2011mp,Hisano:2006mv,Hsieh:2007wq,Blechman:2008gu,Kribs:2008hq,Choi:2008pi,Plehn:2008ae,Harnik:2008uu,Choi:2008ub,Kribs:2009zy,Belanger:2009wf,Benakli:2009mk,Kumar:2009sf,Chun:2009zx,Benakli:2010gi,Fok:2010vk,DeSimone:2010tf,Choi:2010gc,Choi:2010an,Benakli:2011kz,Heikinheimo:2011fk,Fuks:2012im,Kribs:2012gx,Kumar:2011np,Fok:2012fb}.

We follow a recent framework (dubbed the MRSSM) proposed by \cite{Kribs:2007ac}. We investigate the processes $\mu \to e \gamma$, $\mu \to e$ conversion, and $\mu \to eee$.  A scan over all sensitive parameters with respect to $\mu \to e$ mixing was performed \cite{Fok:2010vk} and it was shown that slepton mixing parameters in the MRSSM could be as large as $O(0.1)$ with bounds from current experiments. Furthermore, the most severe constraint is obtained by combining $\mu \to e \gamma$ and $\mu \to e$ conversion on  $\tilde{e}_R-\tilde{\mu}_R$ mixing.

The sensitivities to $\mu \to e$ conversion and $\mu \to eee$ at Project X will be improved by factors $10^6$ and $10^3-10^4$, respectively, over the sensitivities of current experiments  \cite{projectx}. The exclusion plots assuming the above expected sensitivities plots will be presented in this paper. It will be organized as follow. Section \ref{sec:flav} gives a discussion on the flavor mixing in the MRSSM. The results will be presented in section \ref{sec:res}. Finally, we will discuss the results in \ref{sec:diss}.

\section{$\mu \to e$ Mixing in the MRSSM}
\label{sec:flav}
Consider $\tilde{\mu}-\tilde{e}$ mixing in the MRSSM, the mass eigenstates of the two sleptons, $\tilde{l}_i$ can be written as

\begin{equation}
\left(
\begin{array}{c}
\tilde{l}_{1} \\
\tilde{l}_{2}
\end{array}
\right)_{L,R} =
\left(
\begin{array}{cc}
  \cos\theta_{\tilde{l}}& \sin\theta_{\tilde{l}} \\
 -\sin\theta_{\tilde{l}} &  \cos\theta_{\tilde{l}}
\end{array}
\right)_{L,R}
\left(
\begin{array}{c}
\tilde{e} \\
\tilde{\mu}
\end{array}
\right)_{L,R}.
\label{sleptonmixing}
\end{equation}

To understand the dependence of $\mu \to e$ amplitudes on the mixing parameters, consider the $e\tilde{e}N$ vertex, where $l = (e,\mu)$ and $N$ is the neutralino. Dropping the subscripts $L,R$ and upon writing the slepton in the eigenbasis according to \ref{sleptonmixing}, one sees that the interaction is proportional to either $\cos\theta_{\tilde{l}}$ or $\sin\theta_{\tilde{l}}$. Now consider diagrams corresponding to $\mu \to e \gamma$ and $\mu \to e$ conversion where only one slepton runs in the loop. Because in all these diagrams the internal slepton line connects with both the external $\mu$ and $e$ line, the amplitude of all diagrams must be proportional to $\cos\theta_{\tilde{l}} \sin\theta_{\tilde{l}}$ for each slepton ``chirality''. For $\mu \to eee$, the diagrams contain terms that are proportional to $\cos^3\theta_{\tilde{l}} \sin\theta_{\tilde{l}}$ and others that are proportional to $\cos\theta_{\tilde{l}} \sin^3\theta_{\tilde{l}}$ due to the two sleptons running in the loop. Because of the large intensity at Project X, the mixing angle $ \theta_{\tilde{l}}$ can be assumed to be small and the diagrams proportional to $\cos\theta_{\tilde{l}} \sin^3\theta_{\tilde{l}}$ drop out. In other words, for a set given masses, the branching fraction of $\mu \to e \gamma$, $\mu \to e$ conversion and $\mu \to eee$ are all proportional to $\sin^2 2\theta_{\tilde{l}} $. This fact is useful when one tries to estimate the bound on slepton mixing parameters during early runs of Project X.

Focusing on the first two generations of lepton mixing, the parameters sensitive to $\mu \to e$ processes are the mixing parameters $\sin 2\theta_{L, R}$, the bino mass, $M_B$, slepton masses, $m_{1,2}$, down type Higgsino mass $\mu_d$, and the slepton mixing angles $\sin 2\theta_{L,R}$ \cite{Fok:2010vk}.

\begin{figure*}[t!]
\includegraphics[width=0.48\textwidth]{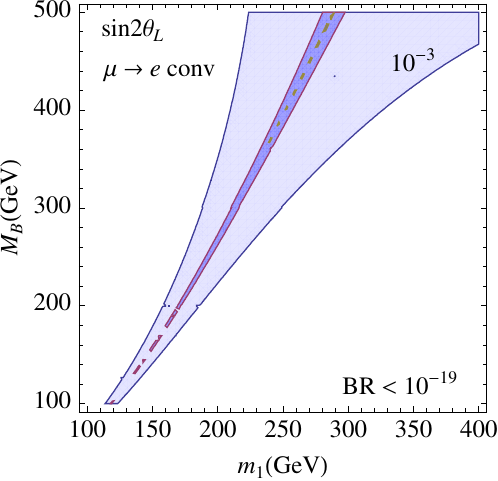}
\hfill
\includegraphics[width=0.48\textwidth]{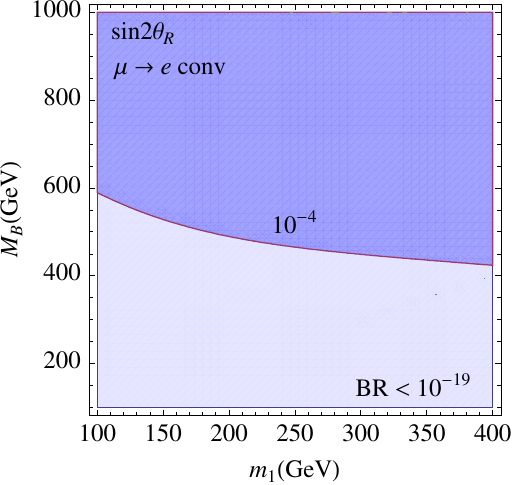}
\caption{$\mu \to e$ conversion exclusion plots with expected limit of $BR(\mu \to e) < 10^{-19}$ at Project X, where $m_1$ is the lightest slepton mass, and $M_B$ is the bino mass, with the slepton masses $m_2 = 1.5 m_1$. The Higgsino mass is $\mu_d = 200$ GeV. The labeled contours are various values of $\sin 2 \theta_{L,R}$. Note that there is significant destructive interfernece among $\mu \to e$ conversion amplitudes a narrow region for the case of left-hand slepton mixing. In the production of these plots the chargino (with mass $\sim O(1$ TeV)) loops are omitted, and the results for left-handed mixing with bino mass up to only 500 GeV is presented for consistency.}
\label{fig:conv}
\end{figure*}

\begin{figure*}[t!]
\includegraphics[width=0.48\textwidth]{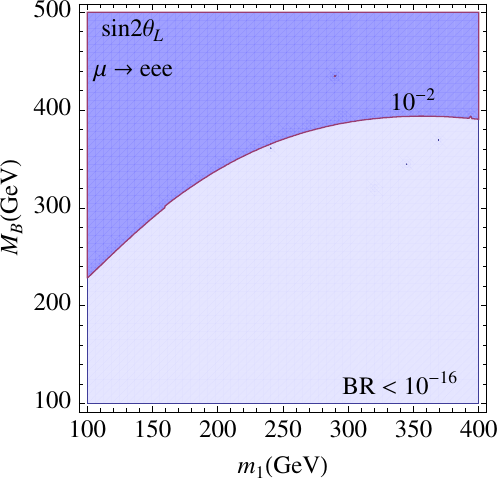}
\hfill
\includegraphics[width=0.48\textwidth]{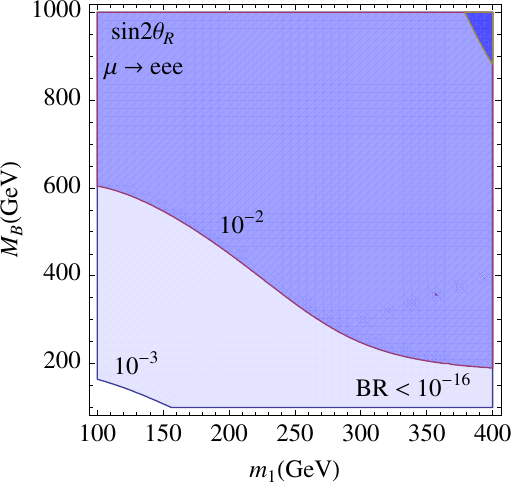}
\caption{Same as FIgure \ref{fig:conv} but for the process $\mu \to eee$.}
\label{fig:eee}
\end{figure*}

\section*{Improvements at Project X}
\label{sec:res}

The exclusion plots of $\mu \to e$ conversion and $\mu \to eee$ with the expected limit at Project X are shown in Figures \ref{fig:conv} and \ref{fig:eee}. The following assumptions have been made to simplify the parameter space. First, the slepton hierarchy is set to have order one splitting. In particular, $m_2 / m_1 = 1.5$, so that the $\mu \to e$ amplitudes are not overly suppressed by the super-JIM mechanism. Second, we only allow either the left handed or right handed mixing to be nonzero, but not both. This is done so that the contribution from each of the left-right sector is manifest. Finally, we have set the down-type Higgsino mass $\mu_d = 200$ GeV and the squark masses $m_{\tilde{q}} = 1$ TeV, where the squark masses appear only in the box diagram of $\mu \to e$ conversion. We also note that the wino mass must be at least $O(1$ TeV) to be consistent with electroweak precision data \cite{Kribs:2007ac}.

The sensitivities to $\mu \to e$ conversion and $\mu \to eee$ are expected to improve by $10^6$ and $10^3$ to $10^4$ respectively. In the scenario where no events from $\mu \to e$ processes are observed, then one would expect the bound on the mixing parameters to be more restricted by $10^{-3}$ for $\mu \to e$ conversion and $10^{-2}$ for $\mu \to eee$. This can be seen in Figures \ref{fig:conv} and \ref{fig:eee}.

\section*{discussion}
\label{sec:diss}

The results show that Project X will be able to constrain the parameter space of $\mu \to e$ mixing in the MRSSM by $\mu \to e$ conversion alone, giving bounds on the mixing parameters  of $\sin 2\theta_{L,R} \lesssim O(10^{-3} - 10^{-4})$ for moderate slepton and bino masses. In the scenario where no $\mu \to e$ events are observed, the non-observation would motivate additional model building to explain the hierarchy between flavor conserving and flavor violating elements of the slepton mass matrix. The results presented here illustrate $BR(\mu \to e) \propto \sin^2 2\theta_{L,R}$. For example, the contours in Figure \ref{fig:eee} are almost identical to the corresponding contours in \cite{Fok:2010vk}, except that the values of $\sin 2\theta_{L,R}$ are scaled down by $10^{-2}$. Also, $\mu \to eee$ provides minimal improvement on the bound on left-handed $\mu \to e$ mixing - it merely excludes the bottom left tail of the incoherence region for $\mu \to e$ conversion. Finally, we checked the bound from $\mu \to e \gamma$ is weaker than the bound obtained in the other two processes in the region of parameter space of interest.

Note that all the masses used in this paper are expected to pass the collider bounds from the LHC. It was shown in \cite{Kribs:2012gx} that the LHC bound on squark masses of the first and second generations in a model with $R$-symmetry is $m_{\tilde{q}} \lesssim 680 - 750$ GeV. Analysis for the bounds on slepton and neutralino masses have not been performed. However, using the squark pair production channels in \cite{Kribs:2012gx} and translating it to slepton pair production, it is safe to expect that the LHC does not give stringent bounds on the slepton masses. The reason is that Drell-Yan is the only production process of the pair production of sleptons at the LHC. Whereas gluon fusion channels contribute to the squark pair production channel. Also, the amplitude of slepton pair production is down by a color factor of 3 and the ratio of electroweak to strong couplings $\alpha_{EW}/ \alpha_s$ when compared to squark pair production. 

\section*{Acknowledgements}
RF would like to thank the organizers for the invitation to  the Project  X workshop in 2012. This work is partially funded by NSERC of Canada.

\begin{appendix}

\label{sec:appa}
%\begin{widetext}
%\end{widetext}

\end{appendix}


\begin{thebibliography}{99}
\begin{widetext}
%\cite{Ahmed:2001eh}
\bibitem{Ahmed:2001eh}
  M.~Ahmed {\it et al.}  [MEGA Collaboration],
  %``Search for the lepton-family-number nonconserving decay \mu -> e +
  %\gamma,''
  Phys.\ Rev.\  D {\bf 65}, 112002 (2002)
  [arXiv:hep-ex/0111030].
  %%CITATION = PHRVA,D65,112002;%%

%\cite{Adam:2009ci}
%\bibitem{Adam:2009ci}
 % J.~Adam {\it et al.}  [MEG collaboration],
  %``A limit for the mu -> e gamma decay from the MEG experiment,''
 % arXiv:0908.2594 [hep-ex].
  %%CITATION = ARXIV:0908.2594;%%

%\cite{Adam:2011ch}
\bibitem{Adam:2011ch} 
  J.~Adam {\it et al.}  [MEG Collaboration],
  %``New limit on the lepton-flavour violating decay $\mu^{+} \to e^{+} \gamma$,''
  Phys.\ Rev.\ Lett.\  {\bf 107}, 171801 (2011)
  [arXiv:1107.5547 [hep-ex]].
  %%CITATION = ARXIV:1107.5547;%%

%\cite{Bertl:2006up}
\bibitem{Bertl:2006up}
  W.~H.~Bertl {\it et al.}  [SINDRUM II Collaboration],
  %``A Search for $\mu - e$ conversion in muonic gold,''
  Eur.\ Phys.\ J.\  C {\bf 47}, 337 (2006).
  %%CITATION = EPHJA,C47,337;%%

%\cite{Bellgardt:1987du}
\bibitem{Bellgardt:1987du}
  U.~Bellgardt {\it et al.}  [SINDRUM Collaboration],
  %``Search For The Decay Mu+ $\ra$ E+ E+ E-,''
  Nucl.\ Phys.\  B {\bf 299}, 1 (1988).
  %%CITATION = NUPHA,B299,1;%%

%\cite{Kutschke:2009zz}
\bibitem{Kutschke:2009zz}
  R.~K.~Kutschke,
  %``The Mu2e experiment at Fermilab,''
  AIP Conf.\ Proc.\  {\bf 1182}, 718 (2009).
  %%CITATION = APCPC,1182,718;%%

\bibitem{projectx}
%\bibitem{projectx}
  Project X and the Science of the Intensity Frontier,
  white paper, Fermilab, 2010. \\
  http://www.fnal.gov/pub/projectx/pdfs/ProjectXwhitepaperJan.v2.pdf


%\cite{Hisano:1995cp}
\bibitem{Hisano:1995cp}
  J.~Hisano, T.~Moroi, K.~Tobe and M.~Yamaguchi,
  %``Lepton-Flavor Violation via Right-Handed Neutrino Yukawa Couplings in
  %Supersymmetric Standard Model,''
  Phys.\ Rev.\  D {\bf 53}, 2442 (1996)
  [arXiv:hep-ph/9510309].
  %%CITATION = PHRVA,D53,2442;%%

%\cite{Hisano:1998fj}
\bibitem{Hisano:1998fj}
  J.~Hisano and D.~Nomura,
  %``Solar and atmospheric neutrino oscillations and lepton flavor violation  in
  %supersymmetric models with the right-handed neutrinos,''
  Phys.\ Rev.\  D {\bf 59}, 116005 (1999)
  [arXiv:hep-ph/9810479].
  %%CITATION = PHRVA,D59,116005;%%

%\cite{Masina:2002mv}
\bibitem{Masina:2002mv}
  I.~Masina and C.~A.~Savoy,
  %``Sleptonarium (constraints on the CP and flavour pattern of scalar lepton
  %masses),''
  Nucl.\ Phys.\  B {\bf 661}, 365 (2003)
  [arXiv:hep-ph/0211283].
  %%CITATION = NUPHA,B661,365;%%

%\cite{Paradisi:2005fk}
\bibitem{Paradisi:2005fk}
  P.~Paradisi,
  %``Constraints on SUSY lepton flavour violation by rare processes,''
  JHEP {\bf 0510}, 006 (2005)
  [arXiv:hep-ph/0505046].
  %%CITATION = JHEPA,0510,006;%%

%\cite{Ciuchini:2007ha}
\bibitem{Ciuchini:2007ha}
  M.~Ciuchini, A.~Masiero, P.~Paradisi, L.~Silvestrini, S.~K.~Vempati and O.~Vives,
  %``Soft SUSY breaking grand unification: Leptons versus quarks on the flavor
  %playground,''
  Nucl.\ Phys.\  B {\bf 783}, 112 (2007)
  [arXiv:hep-ph/0702144].
  %%CITATION = NUPHA,B783,112;%%


%\cite{Fok:2010vk}
%\bibitem{Fok:2010vk} 
 % R.~Fok and G.~D.~Kribs,
  %``\mu to e in R-symmetric Supersymmetry,''
 % Phys.\ Rev.\ D {\bf 82}, 035010 (2010)
  %[arXiv:1004.0556 [hep-ph]].
  %%CITATION = ARXIV:1004.0556;%%
  
 %%%%%%%%%%%%%%%%%%%%%
 
%\cite{Fayet:1978qc}
\bibitem{Fayet:1978qc} 
  P.~Fayet,
  %``Massive Gluinos,''
  Phys.\ Lett.\ B {\bf 78}, 417 (1978).
  %%CITATION = PHLTA,B78,417;%%

%\cite{Polchinski:1982an}
\bibitem{Polchinski:1982an} 
  J.~Polchinski and L.~Susskind,
  %``Breaking of Supersymmetry at Intermediate-Energy,''
  Phys.\ Rev.\ D {\bf 26}, 3661 (1982).
  %%CITATION = PHRVA,D26,3661;%%

%\cite{Hall:1990hq}
\bibitem{Hall:1990hq} 
  L.~J.~Hall and L.~Randall,
  %``U(1)-R symmetric supersymmetry,''
  Nucl.\ Phys.\ B {\bf 352}, 289 (1991).
  %%CITATION = NUPHA,B352,289;%% 
%%%%%%%%%%%%%%%%%%%%%

%\cite{Kribs:2007ac}
\bibitem{Kribs:2007ac}
  G.~D.~Kribs, E.~Poppitz and N.~Weiner,
  %``Flavor in supersymmetry with an extended R-symmetry,''
  Phys.\ Rev.\  D {\bf 78}, 055010 (2008)
  [arXiv:0712.2039 [hep-ph]].
  %%CITATION = PHRVA,D78,055010;%%

%\cite{Fox:2002bu}
\bibitem{Fox:2002bu}
  P.~J.~Fox, A.~E.~Nelson, N.~Weiner,
  %``Dirac gaugino masses and supersoft supersymmetry breaking,''
  JHEP {\bf 0208}, 035 (2002).
  [hep-ph/0206096].


%\cite{Nelson:2002ca}
\bibitem{Nelson:2002ca} 
  A.~E.~Nelson, N.~Rius, V.~Sanz and M.~Unsal,
  %``The Minimal supersymmetric model without a mu term,''
  JHEP {\bf 0208}, 039 (2002)
  [hep-ph/0206102].
  %%CITATION = HEP-PH/0206102;%%

%\cite{Chacko:2004mi}
\bibitem{Chacko:2004mi}
  Z.~Chacko, P.~J.~Fox, H.~Murayama,
  %``Localized supersoft supersymmetry breaking,''
  Nucl.\ Phys.\  {\bf B706}, 53-70 (2005).
  [hep-ph/0406142].

%\cite{Carpenter:2005tz}
\bibitem{Carpenter:2005tz} 
  L.~M.~Carpenter, P.~J.~Fox and D.~E.~Kaplan,
  %``The NMSSM, anomaly mediation and a Dirac bino,''
  hep-ph/0503093.
  %%CITATION = HEP-PH/0503093;%%

%\cite{Antoniadis:2005em}
\bibitem{Antoniadis:2005em} 
  I.~Antoniadis, A.~Delgado, K.~Benakli, M.~Quiros and M.~Tuckmantel,
  %``Splitting extended supersymmetry,''
  Phys.\ Lett.\ B {\bf 634}, 302 (2006)
  [hep-ph/0507192].
  %%CITATION = HEP-PH/0507192;%%

%\cite{Nomura:2005rj}
\bibitem{Nomura:2005rj} 
  Y.~Nomura, D.~Poland and B.~Tweedie,
  %``Minimally fine-tuned supersymmetric standard models with intermediate-scale supersymmetry breaking,''
  Nucl.\ Phys.\ B {\bf 745}, 29 (2006)
  [hep-ph/0509243].
  %%CITATION = HEP-PH/0509243;%%

%\cite{Antoniadis:2006uj}
\bibitem{Antoniadis:2006uj} 
  I.~Antoniadis, K.~Benakli, A.~Delgado and M.~Quiros,
  %``A New gauge mediation theory,''
  Adv.\ Stud.\ Theor.\ Phys.\  {\bf 2}, 645 (2008)
  [hep-ph/0610265].
  %%CITATION = HEP-PH/0610265;%%

%\cite{Kribs:2007ac}
\bibitem{Amigo:2008rc} 
  S.~D.~L.~Amigo, A.~E.~Blechman, P.~J.~Fox and E.~Poppitz,
  %``R-symmetric gauge mediation,''
  JHEP {\bf 0901}, 018 (2009)
  [arXiv:0809.1112 [hep-ph]].
  %%CITATION = ARXIV:0809.1112;%%



%\cite{Benakli:2008pg}
\bibitem{Benakli:2008pg} 
  K.~Benakli and M.~D.~Goodsell,
  %``Dirac Gauginos in General Gauge Mediation,''
  Nucl.\ Phys.\ B {\bf 816}, 185 (2009)
  [arXiv:0811.4409 [hep-ph]].
  %%CITATION = ARXIV:0811.4409;%%

%\cite{Blechman:2009if}
\bibitem{Blechman:2009if} 
  A.~E.~Blechman,
  %``R-symmetric Gauge Mediation and the Minimal R-Symmetric Supersymmetric Standard Model,''
  Mod.\ Phys.\ Lett.\ A {\bf 24}, 633 (2009)
  [arXiv:0903.2822 [hep-ph]].
  %%CITATION = ARXIV:0903.2822;%%

%\cite{Carpenter:2010as}
\bibitem{Carpenter:2010as} 
  L.~M.~Carpenter,
  %``Dirac Gauginos, Negative Supertraces and Gauge Mediation,''
  arXiv:1007.0017 [hep-th].
  %%CITATION = ARXIV:1007.0017;%%

%\cite{Kribs:2010md}
\bibitem{Kribs:2010md} 
  G.~D.~Kribs, T.~Okui and T.~S.~Roy,
  %``Viable Gravity-Mediated Supersymmetry Breaking,''
  Phys.\ Rev.\ D {\bf 82}, 115010 (2010)
  [arXiv:1008.1798 [hep-ph]].
  %%CITATION = ARXIV:1008.1798;%%

%\cite{Abel:2011dc}
\bibitem{Abel:2011dc} 
  S.~Abel and M.~Goodsell,
  %``Easy Dirac Gauginos,''
  JHEP {\bf 1106}, 064 (2011)
  [arXiv:1102.0014 [hep-th]].
  %%CITATION = ARXIV:1102.0014;%%

%\cite{Davies:2011mp}
\bibitem{Davies:2011mp} 
  R.~Davies, J.~March-Russell and M.~McCullough,
  %``A Supersymmetric One Higgs Doublet Model,''
  JHEP {\bf 1104}, 108 (2011)
  [arXiv:1103.1647 [hep-ph]].
  %%CITATION = ARXIV:1103.1647;%%

%\cite{Davies:2011js}
\bibitem{Davies:2011js} 
  R.~Davies and M.~McCullough,
  %``Small neutrino masses due to R-symmetry breaking for a small cosmological constant,''
  Phys.\ Rev.\ D {\bf 86}, 025014 (2012)
  [arXiv:1111.2361 [hep-ph]].
  %%CITATION = ARXIV:1111.2361;%%
  
%\cite{Frugiuele:2011mh}
\bibitem{Frugiuele:2011mh} 
  C.~Frugiuele and T.~Gregoire,
  %``Making the Sneutrino a Higgs with a $U(1)_R$ Lepton Number,''
  Phys.\ Rev.\ D {\bf 85}, 015016 (2012)
  [arXiv:1107.4634 [hep-ph]].
  %%CITATION = ARXIV:1107.4634;%%

%\cite{Itoyama:2011zi}
\bibitem{Itoyama:2011zi} 
  H.~Itoyama and N.~Maru,
  %``D-term Dynamical Supersymmetry Breaking Generating Split N=2 Gaugino Masses of Mixed Majorana-Dirac Type,''
  arXiv:1109.2276 [hep-ph].
  %%CITATION = ARXIV:1109.2276;%%


%%%%%%%%%%%%%%%%%%%%%

%\cite{Hisano:2006mv}
\bibitem{Hisano:2006mv} 
  J.~Hisano, M.~Nagai, T.~Naganawa and M.~Senami,
  %``Electric Dipole Moments in PseudoDirac Gauginos,''
  Phys.\ Lett.\ B {\bf 644}, 256 (2007)
  [hep-ph/0610383].
  %%CITATION = HEP-PH/0610383;%%

%\cite{Hsieh:2007wq}
\bibitem{Hsieh:2007wq} 
  K.~Hsieh,
  %``Pseudo-Dirac bino dark matter,''
  Phys.\ Rev.\ D {\bf 77}, 015004 (2008)
  [arXiv:0708.3970 [hep-ph]].
  %%CITATION = ARXIV:0708.3970;%%

%\cite{Blechman:2008gu}
\bibitem{Blechman:2008gu} 
  A.~E.~Blechman and S.~-P.~Ng,
  %``QCD Corrections to K - anti-K Mixing in R-symmetric Supersymmetric Models,''
  JHEP {\bf 0806}, 043 (2008)
  [arXiv:0803.3811 [hep-ph]].
  %%CITATION = ARXIV:0803.3811;%%

%\cite{Kribs:2008hq}
\bibitem{Kribs:2008hq} 
  G.~D.~Kribs, A.~Martin and T.~S.~Roy,
  %``Supersymmetry with a Chargino NLSP and Gravitino LSP,''
  JHEP {\bf 0901}, 023 (2009)
  [arXiv:0807.4936 [hep-ph]].
  %%CITATION = ARXIV:0807.4936;%%

%\cite{Choi:2008pi}
\bibitem{Choi:2008pi} 
  S.~Y.~Choi, M.~Drees, A.~Freitas and P.~M.~Zerwas,
  %``Testing the Majorana Nature of Gluinos and Neutralinos,''
  Phys.\ Rev.\ D {\bf 78}, 095007 (2008)
  [arXiv:0808.2410 [hep-ph]].
  %%CITATION = ARXIV:0808.2410;%%

%\cite{Plehn:2008ae}
\bibitem{Plehn:2008ae} 
  T.~Plehn and T.~M.~P.~Tait,
  %``Seeking Sgluons,''
  J.\ Phys.\ G G {\bf 36}, 075001 (2009)
  [arXiv:0810.3919 [hep-ph]].
  %%CITATION = ARXIV:0810.3919;%%

%\cite{Harnik:2008uu}
\bibitem{Harnik:2008uu} 
  R.~Harnik and G.~D.~Kribs,
  %``An Effective Theory of Dirac Dark Matter,''
  Phys.\ Rev.\ D {\bf 79}, 095007 (2009)
  [arXiv:0810.5557 [hep-ph]].
  %%CITATION = ARXIV:0810.5557;%%


%\cite{Choi:2008ub}
\bibitem{Choi:2008ub} 
  S.~Y.~Choi, M.~Drees, J.~Kalinowski, J.~M.~Kim, E.~Popenda and P.~M.~Zerwas,
  %``Color-Octet Scalars of N=2 Supersymmetry at the LHC,''
  Phys.\ Lett.\ B {\bf 672}, 246 (2009)
  [arXiv:0812.3586 [hep-ph]].
  %%CITATION = ARXIV:0812.3586;%%

%\cite{Kribs:2009zy}
\bibitem{Kribs:2009zy} 
  G.~D.~Kribs, A.~Martin and T.~S.~Roy,
  %``Squark Flavor Violation at the LHC,''
  JHEP {\bf 0906}, 042 (2009)
  [arXiv:0901.4105 [hep-ph]].
  %%CITATION = ARXIV:0901.4105;%%

%\cite{Belanger:2009wf}
\bibitem{Belanger:2009wf} 
  G.~Belanger, K.~Benakli, M.~Goodsell, C.~Moura and A.~Pukhov,
  %``Dark Matter with Dirac and Majorana Gaugino Masses,''
  JCAP {\bf 0908}, 027 (2009)
  [arXiv:0905.1043 [hep-ph]].
  %%CITATION = ARXIV:0905.1043;%%

%\cite{Benakli:2009mk}
\bibitem{Benakli:2009mk} 
  K.~Benakli and M.~D.~Goodsell,
  %``Dirac Gauginos and Kinetic Mixing,''
  Nucl.\ Phys.\ B {\bf 830}, 315 (2010)
  [arXiv:0909.0017 [hep-ph]].
  %%CITATION = ARXIV:0909.0017;%%

%\cite{Kumar:2009sf}
\bibitem{Kumar:2009sf} 
  A.~Kumar, D.~Tucker-Smith and N.~Weiner,
  %``Neutrino Mass, Sneutrino Dark Matter and Signals of Lepton Flavor Violation in the MRSSM,''
  JHEP {\bf 1009}, 111 (2010)
  [arXiv:0910.2475 [hep-ph]].
  %%CITATION = ARXIV:0910.2475;%%

%\cite{Chun:2009zx}
\bibitem{Chun:2009zx} 
  E.~J.~Chun, J.~-C.~Park and S.~Scopel,
  %``Dirac gaugino as leptophilic dark matter,''
  JCAP {\bf 1002}, 015 (2010)
  [arXiv:0911.5273 [hep-ph]].
  %%CITATION = ARXIV:0911.5273;%%

%\cite{Benakli:2010gi}
\bibitem{Benakli:2010gi} 
  K.~Benakli and M.~D.~Goodsell,
  %``Dirac Gauginos, Gauge Mediation and Unification,''
  Nucl.\ Phys.\ B {\bf 840}, 1 (2010)
  [arXiv:1003.4957 [hep-ph]].
  %%CITATION = ARXIV:1003.4957;%%

%\cite{Fok:2010vk}
\bibitem{Fok:2010vk} 
  R.~Fok and G.~D.~Kribs,
  %``\mu to e in R-symmetric Supersymmetry,''
  Phys.\ Rev.\ D {\bf 82}, 035010 (2010)
  [arXiv:1004.0556 [hep-ph]].
  %%CITATION = ARXIV:1004.0556;%%

%\cite{DeSimone:2010tf}
\bibitem{DeSimone:2010tf} 
  A.~De Simone, V.~Sanz and H.~P.~Sato,
  %``Pseudo-Dirac Dark Matter Leaves a Trace,''
  Phys.\ Rev.\ Lett.\  {\bf 105}, 121802 (2010)
  [arXiv:1004.1567 [hep-ph]].
  %%CITATION = ARXIV:1004.1567;%%

%\cite{Choi:2010gc}
\bibitem{Choi:2010gc} 
  S.~Y.~Choi, D.~Choudhury, A.~Freitas, J.~Kalinowski, J.~M.~Kim and P.~M.~Zerwas,
  %``Dirac Neutralinos and Electroweak Scalar Bosons of N=1/N=2 Hybrid Supersymmetry at Colliders,''
  JHEP {\bf 1008}, 025 (2010)
  [arXiv:1005.0818 [hep-ph]].
  %%CITATION = ARXIV:1005.0818;%%

%\cite{Choi:2010an}
\bibitem{Choi:2010an} 
  S.~Y.~Choi, D.~Choudhury, A.~Freitas, J.~Kalinowski and P.~M.~Zerwas,
  %``The Extended Higgs System in $R$-symmetric Supersymmetry Theories,''
  Phys.\ Lett.\ B {\bf 697}, 215 (2011)
  [Erratum-ibid.\ B {\bf 698}, 457 (2011)]
  [arXiv:1012.2688 [hep-ph]].
  %%CITATION = ARXIV:1012.2688;%%

%\cite{Benakli:2011kz}
\bibitem{Benakli:2011kz} 
  K.~Benakli, M.~D.~Goodsell and A.~-K.~Maier,
  %``Generating mu and Bmu in models with Dirac Gauginos,''
  Nucl.\ Phys.\ B {\bf 851}, 445 (2011)
  [arXiv:1104.2695 [hep-ph]].
  %%CITATION = ARXIV:1104.2695;%%

%\cite{Heikinheimo:2011fk}
\bibitem{Heikinheimo:2011fk} 
  M.~Heikinheimo, M.~Kellerstein and V.~Sanz,
  %``How Many Supersymmetries?,''
  arXiv:1111.4322 [hep-ph].
  %%CITATION = ARXIV:1111.4322;%%

%\cite{Fuks:2012im}
\bibitem{Fuks:2012im} 
  B.~Fuks,
  %``Beyond the Minimal Supersymmetric Standard Model: from theory to phenomenology,''
  arXiv:1202.4769 [hep-ph].
  %%CITATION = ARXIV:1202.4769;%%

%\cite{Kribs:2012gx}
\bibitem{Kribs:2012gx} 
  G.~D.~Kribs and A.~Martin,
  %``Supersoft Supersymmetry is Super-Safe,''
  arXiv:1203.4821 [hep-ph].
  %%CITATION = ARXIV:1203.4821;%%

%\cite{Kumar:2011np}
\bibitem{Kumar:2011np} 
  P.~Kumar and E.~Ponton,
  %``Electroweak Baryogenesis and Dark Matter with an approximate R-symmetry,''
  JHEP {\bf 1111}, 037 (2011)
  [arXiv:1107.1719 [hep-ph]].
  %%CITATION = ARXIV:1107.1719;%%

%\cite{Fok:2012fb}
\bibitem{Fok:2012fb} 
  R.~Fok, G.~D.~Kribs, A.~Martin and Y.~Tsai,
  %``Electroweak Baryogenesis in R-symmetric Supersymmetry,''
  arXiv:1208.2784 [hep-ph].
  %%CITATION = ARXIV:1208.2784;%%
  
\end{widetext}

\end{thebibliography}
\end{document}